\documentclass[aps,pra,amsfonts,amssymb,amsmath,showpacs,
floatfix,nofootinbib,groupedaddress,superscriptaddress,citesort,twocolumn]{revtex4}

\usepackage{amssymb,amsmath,amsthm}

\theoremstyle{definition}

\theoremstyle{remark}

\usepackage{subfigure}
\usepackage{mathrsfs}
\usepackage{longtable}
\usepackage{amsfonts}
\usepackage{amstext}
\usepackage[usenames]{xcolor}
\usepackage[dvips]{graphicx}
\def\qed{\leavevmode\unskip\penalty9999 \hbox{}\nobreak\hfill
     \quad\hbox{\leavevmode  \hbox to.77778em{%
              \hfil\vrule   \vbox to.675em%
               {\hrule width.6em\vfil\hrule}\vrule\hfil}}
     \par\vskip3pt}

\begin{document}
\title{Impossibility of masking a set of quantum states of nonzero measure}

\author{Xiao-Bin Liang}
\email{liangxiaobin2004@126.com}
\affiliation{School of Mathematics and Computer science, Shangrao Normal University, Shangrao 334001, China}
\author{Bo Li}
\email{libobeijing2008@163.com}
\affiliation{School of Mathematics and Computer science, Shangrao Normal University, Shangrao 334001, China}
\author{Shao-Ming Fei}
\email{feishm@cnu.edu.cn}
\affiliation{School of Mathematical Sciences, Capital Normal University, Beijing 100048, China}
\affiliation{Max-Planck-Institute for Mathematics in the Sciences, 04103 Leipzig, Germany}
\author{Heng Fan}
\email{hfan@iphy.ac.cn}
\affiliation{Institute of Physics, Chinese Academy of Sciences, Beijing 100190, China}

\begin{abstract}
We study the quantum information masking based on isometric linear operators that distribute the information encoded in pure states to the correlations in bipartite states.
It is shown that a isometric linear operator can not mask any nonzero measure set of pure states. We present a geometric characterization of the maskable sets, and show that any maskable set must be on a spherical circle in certain Euclidean spaces. Detailed examples and potential applications in such as secret sharing and quantum cryptography are analyzed.
\end{abstract}

\maketitle

\section{Introduction}
The evolution of a closed quantum system is assumed to be unitary in quantum mechanic, which results in some distinguished features like non-cloning theorem \cite{1,2,3}, non-broadcasting theorem \cite{4} and non-deleting theorem \cite{5}. These theorems play an important role in quantum information processing such as key distribution \cite{Hwang,Scarani}, quantum teleportation \cite{Bennett,Bouwmeester} and communication security protocols \cite{Gisin,Samuel}.

In a recent original work \cite{8},  Kavan Modi et. al. studied the problem of quantum information masking. From the unitarity of quantum mechanics, they obtained the so-called no-masking theorem:
it is impossible to mask all arbitrary pure states by the same unitary operator. Different from the decoherence resulted from the interactions between the system and the environment \cite{9,10,11},  quantum masking requires that the information in subsystems are transferred into the correlations of bipartite systems by unitary operations, such that the final reduced states of any subsystems are identical.
Namely, the subsystems themselves contain no longer the initial information.

Different from that non-orthogonal states cannot be perfectly cloned or deleted, there are sets of infinitely many nonorthogonal maskable quantum states \cite{8}. A maskable set may have uncountably many elements that are not orthogonal to each other. Quantum information masking has potential applications in such as secret sharing \cite{M,25,Zhen}. With respect to the deterministic or probabilistic cloning \cite{18,19,20}, deleting and purification \cite{21,22}, there have been a series of results on maskable states,  such as determinate masking schemes in multipartite scenario \cite{Vicente,Li}, probabilistic quantum information masking \cite{Bo}, and the quantum masking machine for states on a spherical circle on the Bloch sphere \cite{Liang}. In \cite{MSL} the authors studied probabilistic and approximate masking of quantum information based on completely positive and trace decreasing (invertible) linear transformations.

In this paper, we study the quantum masking of arbitrary dimensional systems based on isometric linear operations. We claim that no isometric linear operators can mask a nonzero measure set of pure states.
Here, in order to measure the amount of quantum available resources, we give a quantitative  characterization of the maskable sets. In \cite{8} a conjecture has been proposed, which states that the maskable states corresponding to any masker belong to some disk. We show that this is true for some special cases at least. The potential applications of our results in secret sharing are also analyzed.

\section{Isometric linear operators and measure of quantum states}

Let $\mathcal{H}_X$ denote the n-dimensional Hilbert space associated with the system $X$.
We say that an isometric linear operator $\mathcal{U}$ masks the quantum
information contained in a set $\Omega$ of states $\{|a_s\rangle_A \in \Omega\subseteq \mathcal{H}_A\}$, if it maps $|a_s\rangle_A$ to $\{|\Psi_s\rangle_{AB} \in \mathcal{H}_A\otimes \mathcal{H}_B\}$ such that all the marginal states of $|\Psi_s\rangle_{AB}$ are identical:
$\rho_A=\mathrm{Tr}_B(|\Psi_s\rangle_{AB}\langle\Psi_s|)$ and $\rho_B=\mathrm{Tr}_A(|\Psi_s\rangle_{AB}\langle\Psi_s|)$ for all $s$. Namely, the reduced states
$\rho_A$ and $\rho_B$ contain no information about the value of $s$. Here $\Omega$ is said to be the maskable set corresponding to the masker $\mathcal{U}$.

An arbitrary $n$-dimensional pure state $|p\rangle$ can be written as
$$
|p\rangle= \sum^{n}_{k=1} r_{k}e^{{iy_{k-1}}}|k\rangle \equiv|(r_1,...,r_{n},y_1,...,y_{n-1})\rangle,
$$
where $r_k\in[0,1]$, $\sum^{n}_{k=1}r_k^2=1$ and $y_k\in[0,2\pi)$.
For the convenience of discussion, we assume $r_1>0$. Denote $y_0\equiv0$.
We can write $|p\rangle=|(r,y)\rangle$,
where $r=(r_1,...,r_{n})$ and $y=(y_1,...,y_{n-1})$.
Generally, one has the following expressions of $r_1,...,r_{n}$,
\begin{eqnarray}\label{Equation1}
r_1=\cos x_1,~ r_2=\sin x_1\cos x_2,......,\nonumber\\
r_{k}=\sin x_1\sin x_2...\sin x_{k-1}\cos x_{k},~ 2\leq k\leq n-1,...,\nonumber\\
r_{n}=\sin x_1...\sin x_{n-2}\sin x_{n-1},~ x_k\in[0,\frac{\pi}{2}].
\end{eqnarray}
Then $|p\rangle$ can be written as\\ $|p\rangle=|(x_1,...,x_{n-1},y_1,...,y_{n-1})\rangle\equiv|(x,y)\rangle$.

From the domain of the parameters $x_k$ and $y_k$, we can define a ``volume" measure for a set of pure states.  The total volume of all the pure states is $\pi^{2(n-1)}$, i.e., volume measure of the point set $I=[0,\frac{\pi}{2}]^{\times(n-1)}\times[0,2\pi)^{\times(n-1)}$ in the $\mathbb{R}^{2(n-1)}$.
Let $\mathcal{U}$ be an isometric linear operator. For $|p_0\rangle,~|p\rangle\in \mathcal{H}_A$ and $|\Phi_0\rangle,~ |\Phi\rangle \in\mathcal{H}_A\otimes\mathcal{H}_B$ such that
$\mathcal{U}:|p_0\rangle\rightarrow|\Phi_0\rangle$ and $|p\rangle\rightarrow|\Phi\rangle$, we denote
\begin{eqnarray}\label{Equation2}
\Omega_{\mathcal{U}}(|p_0\rangle)=\{|p\rangle:\mathrm{Tr}_A|\Phi\rangle\langle\Phi|
=\mathrm{Tr}_A|\Phi_0\rangle\langle\Phi_0|,\nonumber\\ ~\mathrm{and}~
\mathrm{Tr}_B|\Phi\rangle\langle\Phi|
=\mathrm{Tr}_B|\Phi_0\rangle \langle \Phi_0|\}.
\end{eqnarray}
We say the set $\Omega_{\mathcal{U}}(|(p_0)\rangle)$ is the largest collections of the maskable states with respect to $|p_0\rangle$ and the isometric linear operator $\mathcal{U}$, namely,
the set $\Omega_{\mathcal{U}}(|p_0\rangle)$ is the maskable set with respect to $|p_0\rangle$ and the operator $\mathcal{U}$.
For $|p_0\rangle=|(x^0_1,...,x^0_{n-1},y^0_1,...,y^0_{n-1})\rangle\in I$, the set $\Omega_{\mathcal{U}}(|p_0\rangle)$ can be regarded as a subset of $I\subseteq \mathbb{R}^{2(n-1)}$.
We will show that the ``volume" measure of the set of all maskable states is zero in $\mathbb{R}^{2(n-1)}$.

Without loss of generality, suppose that the isometric linear operator $\mathcal{U}$ acts on the base $|k\rangle$ in the following way,
\begin{eqnarray}
& |k\rangle \rightarrow |\Psi_k\rangle=\sum^{n}_{j=1}|j\rangle|u_{kj}\rangle,~ k=1,\ldots,n,\nonumber
\end{eqnarray}
where
\begin{eqnarray}\label{Equation3}
& |u_{kj}\rangle = \sum^{n}_{m=1}a^m_{kj}|m\rangle,
\end{eqnarray}
and $a^m_{jk} \in\mathbb{C}$, $j,k =1,\ldots,n$.
For an arbitrary pure state $|p\rangle=\sum^{n}_{k=1} r_{k}e^{iy_{k-1}}|k\rangle$, we have
\begin{eqnarray}\label{Equation4}
\mathcal{U}|p\rangle=|\Psi\rangle=\sum^{n}_{k=1} r_{k}e^{iy_{k-1}}\sum^{n}_{j=1}|j\rangle|u_{kj}\rangle.
\end{eqnarray}
The reduced density matrix $\rho_A=\mathrm{Tr}_{B}|\Psi\rangle\langle\Psi|$ is given by
\begin{eqnarray}\label{Equation5}
\rho_A=\sum^{n}_{k=1} \sum^{n}_{l=1} f_{kl}(r_1,...,r_n,y_1,...,y_{n-1})|k\rangle\langle l|,
\end{eqnarray}
where
\begin{eqnarray}\label{fkr}
f_{kl}(p)=f_{kl}(r,y)=\sum^{n}_{j=1} r^2_j  \langle u_{jl}|u_{jk}\rangle \nonumber\\
+\sum^{n}_{t>s} \sum^{n}_{s=1}r_sr_t e^{i(y_{s-1}-y_{t-1})}\langle u_{tl}|u_{sk}\rangle\nonumber\\
+\sum^{n}_{t>s} \sum^{n}_{s=1}r_sr_t e^{i(y_{t-1}-y_{s-1})}\langle u_{sl}|u_{tk}\rangle,
\end{eqnarray}
with $\sum^n_{j}r^2_{j}=1 $ and $y_0=0$.

Now we consider the maskable set $\Omega_{\mathcal{U}}(|(p_0)\rangle)$ in $\mathbb{R}^{2(n-1)}$. According to the definition in Eq. (\ref{Equation2}) for maskable sets, we have $\rho_A=\mathrm{Tr}_A|\Phi\rangle\langle\Phi|
=\mathrm{Tr}_A|\Phi_0\rangle\langle\Phi_0|$, i.e., all the $f_{kl}(p)$ in (\ref{fkr}) should be constant.
Denote $\{P^2_n\}=\{kl:k=1,\cdots,n;\,l=1,\cdots,n\}$. We have $\Omega_{\mathcal{U}}(|(p_0)\rangle)=\{p:f_{kl}(p)\equiv f_{kl}(p_0),kl\in\{P^2_n\} \} $.

That $f_{kl}(p)$ in (\ref{fkr}) are constant implies that both the real part Re$(f_{kl}(p))$ and the imaginary part Im$(f_{kl}(p))$ of $f_{kl}(p)$ are constant functions.
Let $\chi(f)$ denote either the real part Re$(f)$ or the imaginary part Im$(f)$ of $f$. Obviously,
a necessary condition for a maskable set is that $f_{kl}(p)\equiv c_{kl}$ for some complex constants $c_{kl}$.

We first give the following theorem (see proof in Appendix):

\emph{Theorem 1.} No isometric linear operator can mask a set of pure states with nonzero measure.

From Theorem 1 we have immediately the following conclusion.

\emph{Corollary 1.} No isometric masker $\mathcal{U}$ can mask all the pure states.


 \emph{Remark 1.} Corollary 1 is an important result in \cite{8}. Interestingly, Corollary 1 also
implies Theorem 1, namely, Theorem 1 could be deduced from Corollary 1 by some derivations. For more detail, consider a given subset which is maskable if and only if the functions $f_{kl}(p)$ of Eq. (6) for $\rho_A$, and similarly for $\rho_B$, are constant. Obviously, for a masker $\mathcal{U}$, if all $f_{kl}(p)$ are constant for all $p\in I$, it means that masking all quantum states is possible, which contradicts the result of \cite{8}. Otherwise, for the set $\Omega_{\mathcal{U}}(|(p_0)\rangle)=\{p:f_{kl}(p)\equiv f_{kl}(p_0),\,kl\in\{P^2_n\} \} $, there exists a nonzero real function $F(p)=\chi(f_{kl}(p)-f_{kl}(p_0))$,
it's easy to know that $F(p)$ is a real continuous  function,
then by using the well known result in the mathematical literature that the zero set of a real analytical function has Lebesgue measure zero (except for the zero function) \cite{HF}, one can deduce the Theorem 1 immediately. However, the above derivation is qualitative, for better seeking the maskable sets or the isometry maskers and studying the unilateral mask conditions, we provide another method on direct structural proof of Theorem 1 in Appendix.

Theorem 1 gives a quantitative characterization of maskable states. Although with respect to a given isometric masker, the measure of the maskable states is zero, the maskable states may be still infinitely many and uncountable\cite{Liang}. Since the reduced states (local information) are the same for all bipartite states from a maskable set, such quantum masking can be applied to quantum information processing such as secret sharing and quantum cryptography \cite{M,25,Zhen,8}.

\section{ Geometric characteristics of the maskable  quantum state set }

The intersection of spheres and  nontrivial hyperplanes in an n-dimensional Euclid space is called spherical circles $S_n$, namely,
\begin{eqnarray}\label{Equation29}
 S_n=\{(\xi_1,...,\xi_n): \sum (\xi_i-a_i)^2=r^2,\nonumber\\ \sum A_i\xi_i=B, a_i, A_i, B\in\mathbb{R} \mathrm{~and ~not~ all~ be~ zero} \}.\end{eqnarray}
The following theorem shows that a maskable state set is always on a spherical circle in some space.

\emph{Theorem 2.}  The maskable state set is on a spherical circle in a $(P^2_n+n)$-dimensional Euclidean space.

\emph{Proof.}  From (\ref{fkr}), $\chi [f_{kl}(r,y)]= \chi c_{kl}$ gives rise to
\begin{eqnarray}\label{Equation30}
\chi [f_{kl}(r,y)]=\sum^{n}_{j=1} r^2_j A^{kl}_j \nonumber\\
+\sum^{n}_{t>s} \sum^{n}_{s=1}[B^{kl}_{st} \sqrt{2}r_sr_t \cos(y_{s-1}-y_{t-1})+\nonumber\\C^{kl}_{st} \sqrt{2}r_sr_t \sin(y_{s-1}-y_{t-1})]
+D^{kl}=0,\end{eqnarray}
where  $A^{kl}_j, B^{kl}_{st}, C^{kl}_{st}, D^{kl} \in \mathbb{R}$. By proof of Theorem 1,  $A^{kl}_j$ are not all equal and $B^{kl}_{st}$, $C^{kl}_{st}$ can not be all zero, otherwise, it can be deduced that masker is a zero operator.
Denote $\xi_j=(r_j)^2$, $\xi_{st}=\sqrt{2}r_sr_t\cos(y_{s-1}-y_{t-1}),\\ \xi_{ts}=\sqrt{2}r_sr_t\sin(y_{s-1}-y_{t-1})$ with $t>s$. Then
\begin{eqnarray}\label{Equation31}
  \sum^n_{j=1} (\xi_j)^2+\sum_{t>s\geq1 }(\xi_{st})^2\nonumber+\sum_{t>s\geq1 }(\xi_{ts})^2\nonumber\\
  =\sum^n_{j=1} (r_j)^4+ 2\sum_{t>s\geq1 }r^2_sr^2_t\nonumber\\
  =\sum^n_{j=1} (r_j)^4+ \sum^n_{j=1}[(r_j)^2-(r_j)^4]=1.\end{eqnarray}
In other words, a maskable state set satisfies
\begin{eqnarray}\label{Equation32}
 \sum^N_{j=1} (\xi_j)^2=1 ~\mathrm{and}~  \sum^N_{j=1} A_j\xi_j+D=0,
  \end{eqnarray}
where $N=P^2_n+n$, $P^2_n$ is number of all different permutations of 2 different elements taken from n different elements. Obviously, by Theorem 1, hyperplanes not all be trivial, hence a maskable state set is on a spherical circle.  $\blacksquare$

\emph{Remark 2} For qubit states ($N=4$), since $\xi_j=(r_j)^2,~j=1,2$, and $(r_1)^2+(r_2)^2=1$, substituting $\xi_2=1-\xi_1$ into (\ref{Equation32}), we get
\begin{eqnarray}\label{Equation34}
4(\xi_1-\frac{1}{2})^2+2(\xi_3)^2+2(\xi_4)^2=1.
\end{eqnarray}
Denote $\eta_1=2(\xi_1-\frac{1}{2})$, $\eta_2=\sqrt{2}\xi_3$ and $\eta_3=\sqrt{2}\xi_4$.
Eq. (\ref{Equation34}) gives rise to $\sum^3_{j=1} (\eta_j)^2=1$.
By the definition (\ref{Equation29}), $(\eta_1,\eta_2,\eta_3)$ is  a point on a spherical circle in $\mathbb{R}^3$, in fact,  $\eta_1=\cos 2x_1, \eta_2=\sin 2x_1\cos y_1 ,\eta_2=\sin 2x_1\cos y_1$, which corresponds to the pure state on the Bloch sphere. Notice that the plane is not trivial, therefore, the maskable qubit states
corresponding to a linear masker belong to a spherical circle on $\mathbb{R}^3$.
From \cite{Liang}, it is possible to construct an isometry operator to mask
the spherical circle sets of states. However, when the dimension of
$H_A$ is greater than three, the pure states have no equivalent description
similar to the Bloch sphere. Since the pure states $|p\rangle=|(x,y)\rangle$
in $n$-dimensional space $H_A$ has $2(n-1)$ variables, for $n\geq3$ one has no
intuitive picture in general.

In the following, we consider a class of isometry operator maskers and their applications.
First of all, let us consider a maskable set passing through the point $|p_0\rangle$ in 3-dimensional space $H_A$, such that $p_0=(x^0_1,x^0_2,y^0_1,y^0_2)$ and $p=(x_1,x_2,y_1,y_2)$ with $x_1=x^0_1$.
We provide an example of a maskable set,
$$
\begin{array}{l}\Omega(p_0)=\{|p\rangle=|(x_1,x_2,y_1,y_2)\rangle: \cos x_1=\cos x^0_1\equiv a, \nonumber\\
\sin (2x_2)\cos(y_1-y_2)=\sin(2x^0_2)\cos(y^0_1-y^0_2)\equiv b \}.
\end{array}
$$
The maskable set $\Omega(p_0)$ can be masked by the following $\mathcal{U}$:
$$
\begin{array}{l}
 \mid1\rangle\rightarrow|11\rangle,\\[1mm]
 \mid 2\rangle\rightarrow \frac{\sqrt{2}}{2}|22\rangle-\frac{\sqrt{2}}{2}|33\rangle,\\[1mm]
 \mid3\rangle\rightarrow \frac{\sqrt{2}}{2}|22\rangle+\frac{\sqrt{2}}{2}|33\rangle.
 \end{array}
$$
It is easy to verify that the reduced density matrices $\rho_{A,B}=\mathrm{Tr}_{B,A}|\mathcal{U}(p)\rangle\langle\mathcal{U}(p)|$ are given by
$$
\begin{array}{rcl}
\rho_A=\rho_B&=&a^2|1\rangle\langle1|+ \frac{1}{2}(1-a^2)(1+b)|2\rangle\langle2|\\[1mm]
&&+\frac{1}{2}(1-a^2)(1-b)|3\rangle\langle3|.
\end{array}
$$

Set $x_1=x^0_1=0$. For the maskable set $\Omega(p_0)$ passing through the point $p_0=(0,\frac{\pi}{6},\frac{2\pi}{3},\frac{\pi}{4})$, we have the geometrical depiction of
the maskable states, see Fig 1.
\begin{figure}[!htbp]
\includegraphics[scale=0.6]{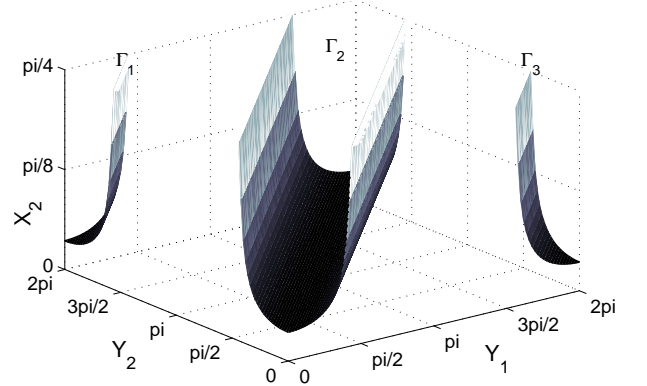}
\caption{The maskable sets $\Omega(p_0)$ passing through the point $p_0$ (associated with the state $|p_0\rangle$), $\Gamma_1$, $\Gamma_2$ and $\Gamma_3$, where $p_0=(0,\pi/6,2\pi/3,\pi/4)$, $\Omega(p_0)=\{p: p=(0,x_2,y_1,y_2)\}$.}\label{fig_1}
\end{figure}

As another example of maskable state set in a 4-dimensional space $H_A$, we consider
$(r_1)^2+(r_2)^2={1}/{2}$ and $(r_3)^2+(r_4)^2={1}/{2}$ for
any $p=|(r_1,r_2,r_3,r_4,y_1,y_2,y_3)\rangle$.
Denote $r_1=\frac{\sqrt{2}}{2}\cos \zeta_1$ and $r_3=\frac{\sqrt{2}}{2}\cos \zeta_2$. Then correspondingly $r_2=\frac{\sqrt{2}}{2}\sin \zeta_1$ and $r_4=\frac{\sqrt{2}}{2}\sin \zeta_2$,
and $p$ can be denoted as $p=(\zeta_1,\zeta_2,y_1,y_2,y_3)$. Let $p_0=(\zeta^0_1,\zeta^0_2,y^0_1,y^0_2,y^0_3)$. We have the following maskable set,
$$
\begin{array}{l}\tilde{\Omega}(p_0)=\{|p\rangle:
   \sin 2\zeta_1\cos y_1= \sin 2\zeta^0_1\cos y^0_1\equiv c,\nonumber\\
 \frac{1}{2}\cos2\zeta_2-\frac{\sqrt{3}}{2}\sin2\zeta_2\cos (y_3-y_2)=\nonumber\\ \frac{1}{2}\cos2\zeta^0_2-\frac{\sqrt{3}}{2}\sin 2\zeta^0_2\cos (y^0_3-y^0_2)\equiv d\}
\end{array}.
$$
The set $\tilde{\Omega}(p_0)$ can be masked by a masker $\tilde{\mathcal{U}}$ such that:
$$
\begin{array}{l}
 \mid1\rangle\rightarrow \frac{\sqrt{2}}{2}|11\rangle-\frac{\sqrt{2}}{2}|22\rangle,~~~
 \mid2\rangle\rightarrow \frac{\sqrt{2}}{2}|11\rangle+\frac{\sqrt{2}}{2}|22\rangle,\\
  \mid3\rangle\rightarrow \frac{\sqrt{6}}{4}e^\frac{i\pi}{4}|3\rangle(|3\rangle+|4\rangle)+\frac{\sqrt{2}}{4}e^\frac{-i\pi}{4}|4\rangle(|3\rangle-|4\rangle),\\
  \mid4\rangle\rightarrow \frac{-\sqrt{2}}{4}e^\frac{i\pi}{4}|3\rangle(|3\rangle+|4\rangle)+\frac{\sqrt{6}}{4}e^\frac{-i\pi}{4}|4\rangle(|3\rangle-|4\rangle).
\end{array}
$$

The reduced  matrices $\rho_{A,B}=\mathrm{Tr}_{B,A}|\tilde{\mathcal{U}}(p)\rangle\langle\tilde{\mathcal{U}}(p)|$ are given by
$$
\begin{array}{rcl}
\rho_A&=&\frac{1}{4}(1+c)|1\rangle\langle1|+ \frac{1}{4}(1-c)|2\rangle\langle2|\\[1mm]
&&+\frac{1}{4}(1+d)|3\rangle\langle3|+\frac{1}{4}(1-d)|4\rangle\langle4|,\\[1mm]
\rho_B&=&\frac{1}{4}\sum^4_{j=1}|j\rangle\langle j|
+ \frac{c}{4}(|1\rangle\langle2|+|2\rangle\langle1|)\\[1mm]
&&+\frac{d}{4}(|3\rangle\langle4|+ |4\rangle\langle3|).
\end{array}
$$
The maskable set $\tilde{\Omega}(p_0)$ passing through the point $p_0$, $p\in \tilde{\Omega}(p_0)$, can be expressed by a two-dimensional graph and a three-dimensional graph, a plane given by $(y_1,\zeta_1)$ and a spatial graph given by $(\zeta_2,y_2,y_3)$, see Fig. 2 for an intuitive description.
\begin{figure*}[htp]
\centering
\subfigure[Line for $ \zeta_1=f(y_1) $ with respect to $y^0_1=\frac{\pi}{6}$ and $\zeta^0_1=\frac{\pi}{4}.$] {\includegraphics[height=2.5in,width=3.0in,angle=0]{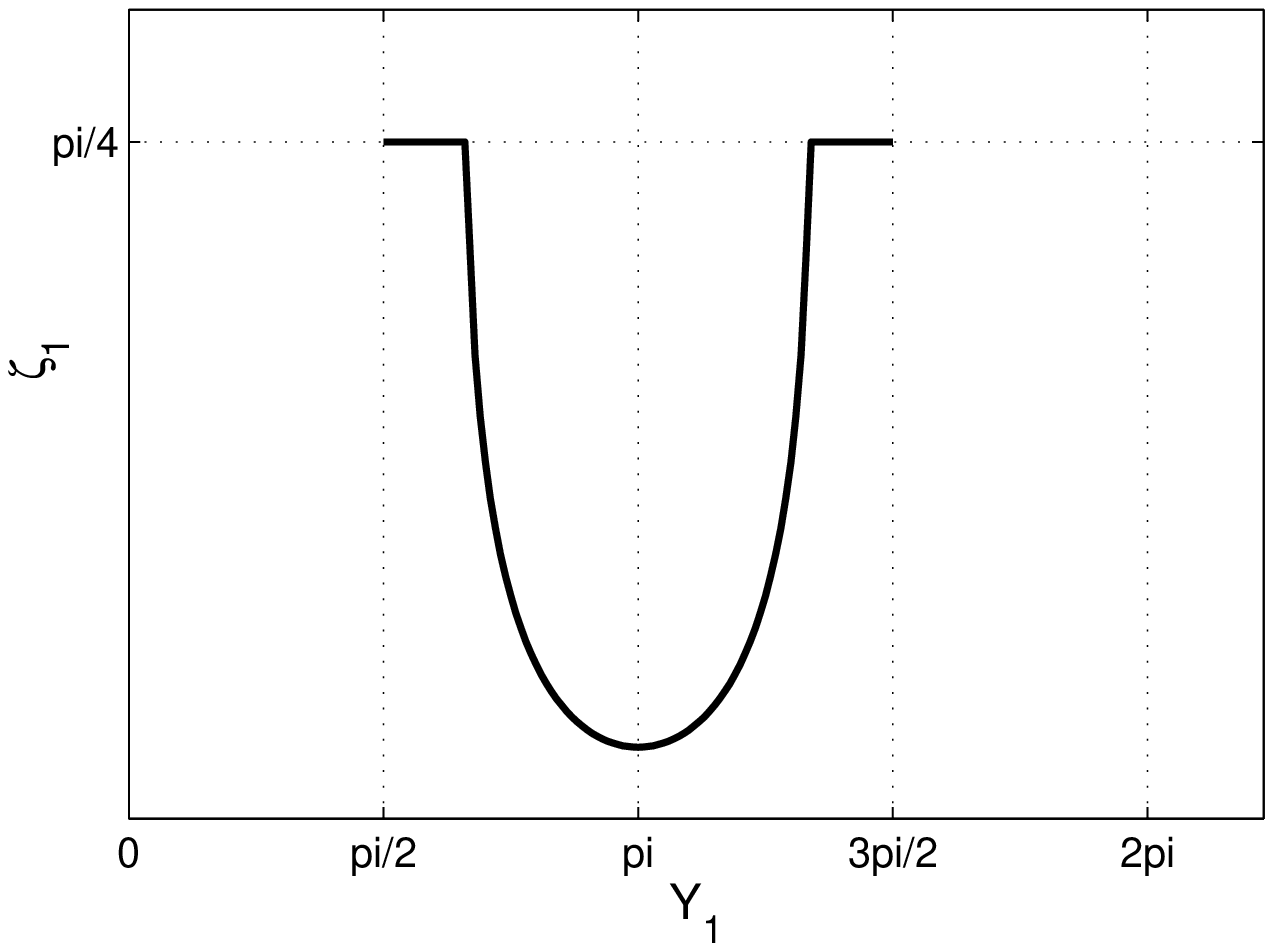}}
\subfigure[Surface for $y_3=F(y_2,\zeta_2)$ with respect to $y^0_2=\frac{2\pi}{3}$, $\zeta^0_2=\frac{2\pi}{3}$ and $y^0_3=\frac{\pi}{4}.$] {\includegraphics[height=2.5in,width=3.0in,angle=0]{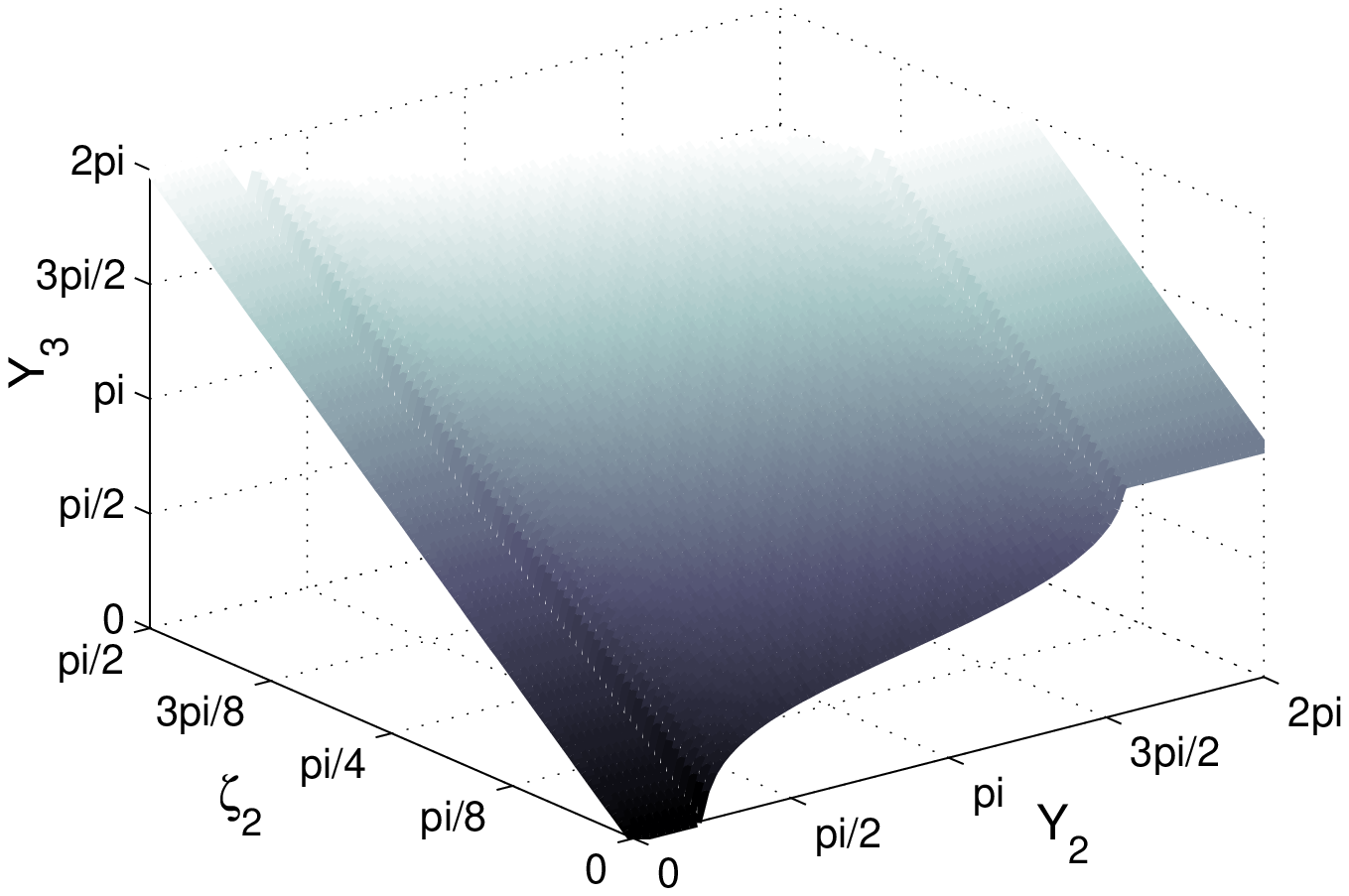}}
\caption{Illustrations of the maskable set $\tilde{\Omega}(p_0)$ passing through the point $p_0$, where $p_0=(\frac{\pi}{4},\frac{2\pi}{3},\frac{\pi}{6}, \frac{\pi}{4},\frac{2\pi}{3})$, $p=(\zeta_1,\zeta_2,y_1,y_2,y_3)$ and $\tilde{\Omega}(p_0)=\{p: \zeta_1=f(y_1),y_3=F(y_2,\zeta_2)\}$.}
\label{fig4}
\end{figure*}

Now we consider the construction of quantum maskers in $n$-dimensional space $H_A$. In \cite{Liang} we have presented explicit maskers for qubit states. We can use these maskers for qubit states to construct maskers for high dimensional states. If $n=2N$ is even,
we write the $n$ dimensional pure states as $|p\rangle=\Sigma^N_{k=1}(s_k|0_k\rangle+t_k|1_k\rangle)$, $s_k,t_k\in\mathbb{C}$, where $|0_k\rangle$ and $|1_k\rangle$, $k=1,..,N$, are the basis of $H_A$.
Namely, we divide the pure state $|p\rangle$ into $N$ ``qubit" parts.
Assume $|s_k|^2 +|t_k|^2=r^2_k$, where $r_k$ is a fixed real value. Then each part is equivalent to a qubit. We can mask  ``the spherical circle sets" of every ``qubit" part by isometry masker $\mathcal{U}_k=\mathcal{S}_{\theta_k}^{\alpha_k}$ given in \cite{Liang}, and $\mathcal{U}_{\theta}^{\alpha}=\sum^N_{k=1} \mathcal{U}_k$ is a quantum masker,
where $\mathcal{U}_k|j_l\rangle=0$ for $k\neq l$, and $\mathcal{U}_k|j_k\rangle= \mathcal{S}_{\theta_k}^{\alpha_k}|j_k\rangle$ for $j=0,1$.
If $n=2N+1$ is odd, we can write the pure states as $|p\rangle=a|1\rangle+\Sigma^N_{k=1}(s_k|0_k\rangle+t_k|1_k\rangle)$.
The corresponding masker is $\mathcal{U}_{\theta}^{\alpha}=U_0+\sum^N_{k=1} \mathcal{U}_k$,
where $U_0|1\rangle=|11\rangle$, $U_0|j_l\rangle=0$ for  $j=0,1$ and $l=1,2,...,N$.

The maskers $\mathcal{U}_{\theta}^{\alpha}$ can used for secret sharing.
Alice encodes the message $p_0$ into the state $|p_0\rangle$.
By applying a masker $\mathcal{U}_{0}^{\alpha}$, she gets bipartite pure states
$|\Psi_{0}^{\alpha}\rangle_{AB}$.
Alice keeps the part $A$, sends the part $B$ to \{$Bob_{0}^{\alpha_1}, Bob_{0}^{\alpha_2},...$\}, and informs them the maskers \{$\mathcal{S}_{0}^{\alpha_1}$, $\mathcal{S}_{0}^{\alpha_2}$,...\}, respectively.
The Bobs can only obtain the information from the reduced states, and cannot decode the information by local quantum operations without classical communication. However, two or more Bobs cooperate together can decode the secret information. Similar potential applications may be considered for quantum cryptography and other quantum communication protocols \cite{27,28}.

\section{Conclusion}

We have presented a quantitative characterization of maskable state sets.
We have shown that isometric linear operators can not mask nonzero measure set of pure states.

Moreover, it has been shown that the maskable quantum state set is on a spherical circle in certain space,
which gives a geometrical characterization the maskable sets.
It should be noted here that although Theorem 2 gives a positive answer to the maskable state
``disk" conjecture raised in \cite{8} in a certain sense , the converse of Theorem 2 does not necessarily hold unless the dimension of $H_A$ is 2.

Furthermore, a class of explicit isometry quantum maskers have been constructed in terms of qubit maskers. These maskers can applied for information processing like secret sharing.
Our results may high light further researches on quantum information masking and its practical applications in quantum communication protocols.

\bigskip
\noindent {\bf Acknowledgments}
We thank the anonymous referees for their helpful suggestions. This work is supported by NSFC under Nos 11765016 and 11675113, Beijing Municipal Commission of Education (KZ201810028042),  Scientific research
project of Jiangxi Provincial Department of Education
(GJJ190888), and Beijing Natural Science Foundation (Z190005).
Xiao-Bin Liang and Bo Li contribute equally to this work.

\section{Appendix}\label{Appendix}

In this appendix, we provide a detail proof of Theorem 1 of the main text.
We first introduce the following Lemma.

\emph{Lemma.} Let $y=f(t)=f(t_1,t_2,...,t_n)$ be a real continuous  function on $E=[a_1,b_1]\times\cdot\cdot\cdot\times[a_n,b_n]\subseteq\mathbb{R}^n$. Denote $R^{n+1}_E(f)=\{(y,t_1,...,t_n)\in \mathbb{R}^{n+1} : y=f(t), t\in E\}$.  We call $R^{n+1}_E(f)$ the graph of function $f$ in $\mathbb{R}^{n+1}$. For any graph of function $f$, we have $M[R^{n+1}_E(f)]=0$, where $M[\cdot]$ is the Lebesgue measure. Namely, the measure of the graph of $y=f(t)$ is zero in $\mathbb{R}^{n+1}$\cite{V}.

Since $f_{kl}(p)\equiv c_{kl}$ if and only if $ \chi [f_{kl}(p)]\equiv \chi [c_{kl}]$, we have $\Omega_{\mathcal{U}}(|(p_0)\rangle)=\{p:\chi [f_{kl}(p)]\equiv \chi [c_{kl}],kl\in\{P^2_n\} \}$.
We use Lemma  to evaluate the measure of a maskable set. The condition $\chi [f_{kl}(p)]\equiv \chi [c_{kl}]$ can be reduced to the following forms:
\begin{eqnarray}\label{Equation9}
\chi[f_{kl}(r_1,\ldots,r_n,y_1,...,y_{n-1})]=\chi\hbar_{kl}(r)+ ~~~~~~\nonumber\\
\chi\mu^1_{kl}(r,y_2,...,y_{n-1})\cos y_1+\chi\nu^1_{kl}(r,y_2,...,y_{n-1})\sin y_1+\nonumber\\
\chi\mu^2_{kl}(r,y_3,...,y_{n-1})\cos y_2+\chi\nu^2_{kl}(r,y_3,...,y_{n-1})\sin y_2+ \nonumber\\...+
\chi\mu^{n-2}_{kl}(r,y_{n-1})\cos y_{n-2} +\chi\nu^{n-2}_{kl}(r,y_{n-1})\sin y_{n-2}+\nonumber\\
\chi\mu^{n-1}_{kl}(r)\cos y_{n-1} +\chi\nu^{n-1}_{kl}(r)\sin y_{n-1}=\chi(c_{kl}),~~ \end{eqnarray}
where $\chi\hbar_{kl}$, $\chi\mu^j_{kl}$ and $\chi\nu^j_{kl}$ are real functions that can be obtained by separating the real and imaginary parts of (\ref{fkr}), $j=1,...,n-1$. In particular,
\begin{eqnarray} \chi\mu^{n-1}_{kl}(r)=\chi[r_1r_n(\langle u_{1l}|u_{nk}\rangle+i\langle u_{nl}|u_{1k}\rangle)],\nonumber\\ \chi\nu^{n-1}_{kl}(r)=\chi[r_1r_n(\langle u_{1l}|u_{nk}\rangle-i\langle u_{nl}|u_{1k}\rangle)].
\end{eqnarray}
When $r_1, r_n\neq0$, it is easy to see that
\begin{eqnarray}\label{Equation11}
|\mu^{n-1}_{kl}(r)|^2+|\nu^{n-1}_{kl}(r)|^2=0~~\mathrm{if~ and~ only~ if}\nonumber\\
\langle u_{1l}|u_{nk}\rangle=\langle u_{nl}|u_{1k}\rangle=0.
\end{eqnarray}

Accounting to the symmetry of $y_1, ..., y_{n-1}$ in Eq. (\ref{fkr}), we denote $\{y_{\tau(1)},
..., y_{\tau(n-1)}\}$ a permutation of $\{y_1, ..., y_{n-1}\}$. Applying such permutation to (\ref{Equation9}) one obtains a similar result to Eq. (\ref{Equation11}), $|\mu^{\tau(n-1)}_{kl}(r)|^2+|\nu^{\tau(n-1)}_{kl}(r)|^2=0$ if and only if
\begin{eqnarray*}
\langle u_{1l}|u_{\tau(n)k}\rangle=\langle u_{\tau(n)l}|u_{1k}\rangle=0.
\end{eqnarray*}

Denote $E_{\delta}=[\delta,\frac{\pi}{2}-\delta]^{\times(n-1)}\times[\delta,2\pi-\delta]^{\times(n-1)}\subset I$.  Let us consider the measure of set the $E=\Omega_{\mathcal{U}}(|(p_0)\rangle)\bigcap E_{\delta}$.
In order to prove that the measure of $\Omega_{\mathcal{U}}(|(p_0)\rangle)$ is zero in $I\subseteq \mathbb{R}^{2(n-1)}$, one only needs to prove that the measure of $E$ is zero.

\emph{Theorem 1.} No  isometric linear operator can mask a set of pure states with nonzero measure.

\emph{\textbf{Proof}.} For the  operator ${\mathcal{U}} $ defined in Eq. (\ref{Equation3}-\ref{fkr}), we consider the measure of the set $\Omega_{\mathcal{U}}(|(p_0)\rangle)=\{p:\chi[f_{kl}(p)]\equiv \chi[f_{kl}(p_0)],kl\in\{P^2_n\} \} $. From the previous analysis and Lemma, it suffices to verify that the measure of the graph of  Eq. (\ref{Equation9}) is zero in $E_\delta \subseteq \mathbb{R}^{2(n-1)}$. If  $\mid\mu^1_{kl}(\star)\mid^2+\mid\nu^1_{kl}(\star)\mid^2\neq0$, here $\star$ stands for $(r,y_2,...,y_{n-1})$ in Eq. (\ref{Equation9}), then from (9) one easily finds a continuous real solution of $y_1$ of the form $y_1=\arcsin(\omega(r,y_2,...,y_{n-1}))+\phi(r,y_2,...,y_{n-1})$. According to the parameterizations (1) $y_1$ can be generally written as
\begin{eqnarray}\label{Equation12}
y_1=f(x,y_2,...,y_{n-1}),\end{eqnarray}  where $(x,y_2,...,y_{n-1})\in R^{2n-3}$.

Let $\Omega_1=\{(x,y_1,y_2,...,y_{n-1}): y_1=f(x,y_2,...,y_{n-1})\}$. Then  $E=\Omega_{\mathcal{U}}(|(p_0)\rangle)\bigcap E_{\delta}\subseteq\Omega_1\bigcap E_\delta$.
By Lemma, the measure of $\Omega_1\bigcap E_\delta$ in $\mathbb{R}^{2(n-1)}$ is zero, i.e., the measure of $E$ is zero. Letting $\delta\rightarrow0$, we have that the measure of $\Omega_{\mathcal{U}}(|(p_0)\rangle)$ is zero in $I$, which proves the theorem for this case.

Otherwise, if $\mid\mu^1_{kl}(\star)\mid^2+\mid\nu^1_{kl}(\star)\mid^2=0$, then Eq. (\ref{Equation9}) reduces to
\begin{eqnarray}\label{Equation13}
\chi[f_{kl}(r_1,...,r_n,y_2,...,y_{n-1})]=\chi\hbar_{kl}(r)+ ~~~~~~\nonumber\\ \chi\mu^2_{kl}(r,y_3,...,y_{n-1})\cos y_2+\chi\nu^2_{kl}(r,y_3,...,y_{n-1})\sin y_2+ \nonumber\\...+\chi\mu^{n-2}_{kl}(r,y_{n-1})\cos y_{n-2} +\chi\nu^{n-2}_{kl}(r,y_{n-1})\sin y_{n-2}+\nonumber\\
\chi\mu^{n-1}_{kl}(r)\cos y_{n-1} +\chi\nu^{n-1}_{kl}(r)\sin y_{n-1}=\chi(c_{kl}).~~~ \end{eqnarray}
We consider whether $|\mu^2_{kl}(\star)|^2+|\nu^2_{kl}(\star)|^2$ is zero or not, here the notation  $\star$ similarly represents $(r,y_3,...,y_{n-1})$ in Eq. (\ref{Equation9}). If $|\mu^2_{kl}(\star)|^2+|\nu^2_{kl}(\star)|^2\neq 0$, the theorem is proved. Otherwise,
one discusses the case $y_2,y_3,...,y_{n-1}$ in a similar manner. Finally, one may need to consider the only case,
\begin{eqnarray}\label{Equation14} |\mu^{n-1}_{kl}(r)|^2+|\nu^{n-1}_{kl}(r)|^2=0 .\end{eqnarray}
From (\ref{Equation11}), we have
\begin{eqnarray}\label{Equation15}
\langle u_{1l}|u_{nk}\rangle= 0,~\langle u_{nl}|u_{1k}\rangle= 0.
\end{eqnarray}
Due to permutation symmetry of $y_1, ..., y_{n-1}$ in Eq. (\ref{fkr}), we have,
\begin{eqnarray}\label{Equation16}
\langle u_{1l}|u_{sk}\rangle= 0~~ \mathrm{and} ~~\langle u_{sl}|u_{1k}\rangle= 0,~ s=2,...,n-1.
\end{eqnarray}
Substituting (\ref{Equation16}) into (\ref{fkr}), we obtain
\begin{eqnarray}\label{Equation17}
f_{kl}(r,y)=\hbar_{kl}(r)
+\sum^{n}_{t>s} \sum^{n}_{s=2}r_sr_t e^{i(y_{s-1}-y_{t-1})}\langle u_{tl}|u_{sk}\rangle\nonumber\\
+\sum^{n}_{t>s} \sum^{n}_{s=2}r_sr_t e^{i(y_{t-1}-y_{s-1})}\langle u_{sl}|u_{tk}\rangle.~~
\end{eqnarray}

Denote $\tilde{y}_m=y_{m+1}-y_{1}$, $m=1,...,n-2,$ and $\tilde{y}_0\equiv0$.
Equation (\ref{Equation17}) can be rewritten as
 \begin{eqnarray}\label{Equation18}
f_{kl}(x,y)=\hbar_{kl}(r)
+\sum^{n}_{t>s} \sum^{n}_{s=2}r_sr_t e^{i(\tilde{y}_{s-2}-\tilde{y}_{t-2})}\langle u_{tl}|u_{sk}\rangle\nonumber\\
+\sum^{n}_{t>s} \sum^{n}_{s=2}r_sr_t e^{i(\tilde{y}_{t-2}-\tilde{y}_{s-2})}\langle u_{sl}|u_{tk}\rangle.
\end{eqnarray}
Expressing (\ref{Equation18}) in a similar form to (\ref{Equation9}), one has
\begin{eqnarray}\label{Equation19}
\chi[f_{kl}(r,\tilde{y}_1,...,\tilde{y}_{n-2})]={\hbar}_{kl}(r)+ ~~~~~~\nonumber\\
\tilde{\mu}^1_{kl}(r,\tilde{y}_2,...,\tilde{y}_{n-2})\cos \tilde{y}_1+\tilde{\nu}^1_{kl}(r,\tilde{y}_2,...,\tilde{y}_{n-2})\sin \tilde{y}_1\nonumber\\
+ ...+
\tilde{\mu}^{n-1}_{kl}(r)\cos \tilde{y}_{n-2} +\tilde{\nu}^{n-1}_{kl}(r)\sin \tilde{y}_{n-2}=\chi(c_{kl}). \end{eqnarray}
Again if $|\tilde{\mu}^1_{kl}(\star)|^2+|\tilde{\nu}^1_{kl}(\star)|^2\neq0$, then there exist continuous real function $\tilde{y}_{1}=y_2-y_1=\tilde{F}_1(x,\tilde{y}_2,...,\tilde{y}_{n-2})$, namely,
$y_1=f(x,y)$. By Lemma, the measure of $\Omega_{\mathcal{U}}(|(p_0)\rangle)$ is zero in $I$. Otherwise, we continue our discussion for $\tilde{y}_2,\tilde{y}_3,...,\tilde{y}_{n-2}$. Repeating the steps (\ref{Equation13})-(\ref{Equation15}), we get
\begin{eqnarray}\label{Equation20}
\langle u_{2l}|u_{sk}\rangle= 0~~ \mathrm{and} ~~\langle u_{sl}|u_{2k}\rangle= 0,~ s=3,...,n-1.
\end{eqnarray}
Then repeating the steps (\ref{Equation17})-(\ref{Equation20}), one has either the theorem hold or eventually gets
\begin{eqnarray}\label{Equation21}
\langle u_{tl}|u_{sk}\rangle=\langle u_{sl}|u_{tk}\rangle=0,~ s\neq t.
\end{eqnarray}

Therefore, from (\ref{Equation1}) and (\ref{Equation21}), (\ref{fkr}) can be reduced to the following form,
\begin{eqnarray}\label{Equation22}
f_{kl}(r,y)=\sum^{n}_{j=1} r^2_j  \langle u_{jl}|u_{jk}\rangle\nonumber\\
=\langle u_{1l}|u_{1k}\rangle(\cos x_1)^2+\langle u_{2l}|u_{2k}\rangle(\sin x_1\cos x_2)^2\nonumber\\
+\langle u_{3l}|u_{3k}\rangle(\sin x_1\sin x_2\cos x_3)^2+...\nonumber\\
+\langle u_{n-2l}|u_{n-2k}\rangle(\sin x_1...\sin x_{n-2}\cos x_{n-1})^2\nonumber\\
+\langle u_{n-1l}|u_{n-1k}\rangle(\sin x_1...\sin x_{n-2}\sin x_{n-1})^2.
\end{eqnarray}

If $\langle u_{n-2l}|u_{n-2k}\rangle\neq\langle u_{n-1l}|u_{n-1k}\rangle$, then from  $\chi f_{kl}(x,y)=\chi c_{kl}$, there exists a continuous real function $\varphi$ such that $x_{n-1}=\varphi(x_1,x_2,...,x_{n-2})$. By Lemma, we can similarly deduce that the measure of $\Omega_{\mathcal{U}}(|(p_0)\rangle)$ is zero.

If $\langle u_{n-2l}|u_{n-2k}\rangle=\langle u_{n-1l}|u_{n-1k}\rangle$, then we need to discuss $x_{n-2}$,
$x_{n-3}$ and so on. At most, it may end up with
\begin{eqnarray}\label{Equation23}
\langle u_{ml}|u_{mk}\rangle=\langle u_{hl}|u_{hk}\rangle,~ m,h=1,...,n.
\end{eqnarray}
Obviously, $|u_{kj}\rangle$, $kj\in\{P^2_n\}$, can not be all zero. One may assume that
$|u_{11}\rangle\neq0$. By (\ref{Equation23}) then one gets $\||u_{11}\rangle\|^2=\||u_{h1}\rangle\|^2\neq0$,
namely, $|u_{h1}\rangle\neq0~ $, $h=1,2,...,n$. By (\ref{Equation21}), we have $\langle u_{k1}|u_{h1}\rangle=0$ for $h\neq k$.
Therefore, $\{|u_{11}\rangle,|u_{21}\rangle,...,|u_{n1}\rangle\}$ can span the space $H_B$.
Let \begin{eqnarray}\label{Equation24}
|u_{1m}\rangle=\lambda^{1m}_{11}|u_{11}\rangle+\lambda^{1m}_{21}|u_{21}\rangle+\cdots+\lambda^{1m}_{n1}|u_{n1}\rangle.
\end{eqnarray}

If $k\neq1$, by (\ref{Equation21}) and (\ref{Equation24}), we have $ \langle u_{k1}|u_{1m}\rangle=0=\lambda^{1m}_{k1}$. Namely, $|u_{1m}\rangle=\lambda^{1m}_{11}|u_{11}\rangle.$
For the same reason, $|u_{hm}\rangle=\lambda^{hm}_{h1}|u_{h1}\rangle$, $h=1,...,n.$
Since $\||u_{11}\rangle\|^2=\||u_{h1}\rangle\|^2$ and
\begin{eqnarray}
 \langle u_{11}|u_{1m}\rangle=\lambda^{1m}_{11}\||u_{11}\rangle\|^2,~
 \langle u_{h1}|u_{hm}\rangle=\lambda^{hm}_{h1}\||u_{h1}\rangle\|^2,\nonumber
\end{eqnarray}
we get $\lambda^{1m}_{11}=\lambda^{hm}_{h1}=\lambda_m$.

This also means that $|u_{km}\rangle=\lambda_m|u_{k1}\rangle$, $k=1,...,n$. From (\ref{Equation3}), the action of the linear operator $\mathcal{U}$ on the base $|k\rangle$ can also be expressed as
\begin{eqnarray}\label{Equation25}
& |k\rangle \rightarrow |\Psi_k\rangle=(|1\rangle+\lambda_2|2\rangle+...+\lambda_n|n\rangle)|u_{k1}\rangle.
\end{eqnarray}
Denote $|c\rangle=(|1\rangle+\lambda_2|2\rangle+...+\lambda_n|n\rangle)$. The above formula can be transformed into
\begin{eqnarray}\label{Equation26}
& |k\rangle \rightarrow |\Psi_k\rangle=|c\rangle|u_{k1}\rangle=|c\rangle\sum^{n}_{j=1}a^j_{k1}|j\rangle.
\end{eqnarray}
Denote $a^j_{k1}=a_{kj}$. The reduced density matrix $\rho_B$ is given by
\begin{eqnarray}\label{Equation27}
\rho_B=\||c\rangle\|^2(\sum^{n}_{k=1} \sum^{n}_{l=1} g_{kl}(p)|k\rangle\langle l|),
\end{eqnarray}
where
\begin{eqnarray}\label{Equation28}
g_{kl}(r,y)=\sum^{n}_{j=1} r^2_j  \langle a_{jl}|a_{jk}\rangle\nonumber\\
\sum^{n}_{t> s} \sum^{n}_{s=1}r_sr_t e^{i(y_{s-1}-y_{t-1})}\langle a_{tl}|a_{sk}\rangle\nonumber\\
+\sum^{n}_{t> s} \sum^{n}_{s=1}r_sr_t e^{i(y_{t-1}-y_{s-1})}\langle a_{sl}|a_{tk}\rangle.
\end{eqnarray}

Comparing (\ref{fkr}) with (\ref{Equation28}), one may find that they have the same form. Repeating the same analysis as for $\rho_A$, we can draw the conclusions parallel to (\ref{Equation21}) and (\ref{Equation23}), namely, $\langle a_{tl}|a_{sk}\rangle=\langle a_{sl}|a_{tk}\rangle=0$, $s\neq t;$  $ \langle a_{ml}|a_{mk}\rangle=\langle a_{hl}|a_{hk}\rangle$.  Since isometric linear operator  $\mathcal{U}$ is a nonzero operator, one may assume that $a_{11}\neq0$. We have that $a_{11}a_{21}^*=0$ and $|a_{11}|=|a_{21}|$ cannot be true simultaneously. Therefore, for any isometric  operator $\mathcal{U }$, the measure of $\Omega_\mathcal{U}(p_0)$  is zero in $\mathbb{R}^{2(n-1)}$. $\blacksquare$

\end{document}